\documentclass[aps,prl,preprint,groupedaddress]{revtex4}
\newcommand{\ket}[1]{|#1\rangle}

\usepackage{graphicx}
\usepackage{amsmath}

\begin{document}

\title{Spin- and time-resolved photoemission studies of thin $\rm Co_2FeSi$ Heusler alloy films}

\author{J.-P. W\"ustenberg}
\email{jpwuest@physik.uni-kl.de}
\author{M. Cinchetti}
\author{M. S\'{a}nchez Albaneda}
\author{M. Bauer}
\author{M. Aeschlimann}

\affiliation{University of Kaiserslautern, Physics Department,
Erwin-Schr\"{o}dinger-Str.\ 46, 67663 Kaiserslautern, Germany}

\begin{abstract}
We have studied the possibly half metallic $Co_2FeSi$ full Heusler
alloy by means of spin- and time-resolved photoemission
spectroscopy. For excitation, the second and fourth harmonic of
femtosecond Ti:sapphire lasers were used, with photon energies of
$3.1\,eV$ and $5.9\,eV$, respectively. We compare the dependence of
the measured surface spin polarization on the particular
photoemission mechanism, i.e. 1-photon-photoemission (1PPE) or
2-photon photoemission (2PPE). The observed differences in the spin
polarization can be explained by a spin-dependent lifetime effect
occurring in the 2-photon absorption process. The difference in
escape depth of the two methods in this context suggests that the
observed reduction of spin polarization (compared to the bulk)
cannot be attributed just to the outermost surface layer but takes
place at least $4-6\,nm$ away from the surface.
\end{abstract}



\maketitle

The success of modern spintronics devices, i.e. devices relying on
the electron spin as carrier of information depends crucially on the
ability to store, transport and manipulate the spin state of an
electron within a properly chosen material system. While storage is
traditionally accomplished by the use of magnetoresistive effects in
magnetic multilayer systems, the latter problems seem to be
trackable by the use of (partly) semiconductive materials, where
typical spin diffusion lengths are much larger and material
properties can be tuned precisely to establish a coupling to
external fields for coherent spin manipulation purposes
\cite{zutic04}. However, the conductivity mismatch
\cite{Schmidt00,Attema06} between ferromagnetic metals and
semiconductors which hinders efficient spin injection is still an
unsolved issue. A promising class of materials to overcome this
issue consists of half metallic Heusler alloys, exhibiting metallic
behavior for one spin direction and a band gap for the other. This
leads to a full spin polarization $P$ at the Fermi level $E_F$,
which is commonly defined as the normalized difference of spin up
and spin down electron occupation numbers with respect to a certain
energy and given quantization axis:
\begin{equation}
P(E,t)=\frac{n_\uparrow(E,t)-n_\downarrow(E,t)}{n_\uparrow(E,t)+n_\downarrow(E,t)}
\end{equation}
 Fabrication of such alloys with predictable bulk
properties is nowadays routinely possible. This is shown by the
increasing coherence between measured and predicted values for the
element-specific number of magnetic moments $\mu_B$ per unit cell
\cite{Kandpal06,Wurmehl05}. However, the connection to the spin
polarization as the relevant parameter is not straightforward and
relies heavily on the chosen model potential in calculations. In
the case of spin transport between different materials the
interface region is of extreme importance, since various
mechanisms may drastically reduce the spin polarization of electrons
in the surface region. To our knowledge the predicted value of
\mbox{$P=100\,\%$} has not yet been observed in Heusler compounds.
Moreover, the value of the surface spin polarization (SSP) depends
on temperature. Surface sensitive measurements performed on
$Co_2MnSi$ yield values no more than $12\,\%$ at room temperature,
and only at low temperatures ($<20\,K$) values up to $60\,\%$ have
been reported (\cite{Wang05}, and references therein). For the half
Heusler compound $NiMnSb$, Kolev and coworkers excluded surface
states, chemical disorder as well as structural defects as sources
of reduced SSP \cite{Kolev05}. Our experiments with spin-resolved
photoemission techniques indicate that also for thin films of the
full Heusler compound $Co_2FeSi$ (CFS), the spin polarization of
electrons emitted from the surface at $E_F$ is comparable to the
results above. We observe differences in the absolute SSP value,
depending on the employed photo-excitation scheme (1PPE and 2PPE).
With a simple theoretical model we show that these differences can
be explained by the so called spin filter effect
\cite{Aeschlimann97,Aeschlimann98}. We conclude that the SSP loss is
not a single layer effect but extends over a distance of at least
four times the escape depth of standard photoemission, i.e.
$4-6\,nm$.\\

\begin{figure}[t]
\begin{center}
\includegraphics[width=6cm,keepaspectratio=true]{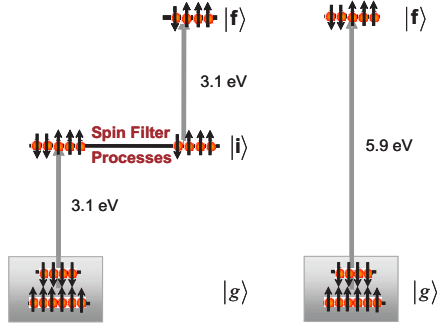}
\caption{Photoemission schemes used in this work. Left panel:
2-Photon absorption (2PPE) within one single pulse. Right panel:
conventional photoemission (1PPE).} \label{pes}
\end{center}
\end{figure}

For our measurements we used both conventional spin-polarized
photoemission spectroscopy (SP-1PPE) as well as spin-polarized
two-photon photoemission spectroscopy (SP-2PPE). The two excitation
schemes are displayed in fig. \ref{pes}.
In a conventional SP-1PPE experiment (right panel in fig.
\ref{pes}), an electron is excited directly from its ground state
$\ket{g}$ below the Fermi energy $E_F$ into an evanescent state
$\ket{f}$ slightly above the vacuum level $E_{vac}$
($\hbar\omega\gtrsim \Phi$, $\Phi=$ work function). After passing
a kinetic energy filter (Focus CSA 300), the longitudinal in-plane
SSP component is analyzed by means of spin-polarized low energy
electron diffraction at a tungsten crystal (Focus SPLEED). The
spectra were recorded at room temperature using the $4^{th}$
harmonic of a pulsed Ti:Sapphire oscillator (Spectra Physics
Tsunami), created by twofold frequency doubling within two thin
\mbox{$\beta$-barium-borate} (BBO) crystals. The average energy
per pulse is $37\,nJ$ at $\hbar \omega=5.9\,eV$. In SP-2PPE, the
excitation is accomplished via two-photon absorption within a
single laser pulse. In this process, the first photon excites an
electron from its ground state $\ket{g}$ below $E_F$ into an
intermediate state $\ket{i}$ below $E_{vac}$. In this state,
dephasing as well as decay processes take place, both being
potentially spin-dependent \cite{Aeschlimann97}. The second photon
is used to excite the electron into the final state $\ket{f}$
known from the 1PPE process. In this experiment we used the
frequency doubled output of a Ti:Sapphire oscillator (Femtosource)
with photon energy $h\nu=3.1\,eV$, pulse width of $40\,fs$ and
$10\,\mu J$ average energy per pulse.  Since the electron in state
$\ket{i}$ may propagate prior to photoemission, the escape depth
of 2PPE is larger than for 1PPE. Thus, we can conclude on the
extension of the spin-depolarizing surface layers by comparing the
spectra obtained by the two methods.

We used a CFS Heusler alloy for our investigations. CFS alloys
possess a high magnetic moment of almost $6\,\mu_B$ per unit cell,
as well as a high Curie temperature of $1100 K$
\cite{Kandpal06,Wurmehl05}. A $70\,nm$ thin film was grown
epitaxially on a $MgO(100)$ substrate. The $4\,nm$ Al cap layer was
removed under UHV conditions \emph{in situ} by
 sputtering with $500\,eV$ $Ar^+$ ions. Reproducible results were
obtained, if the sample was sputtered and annealed (at $570\,K$)
several times prior to the measurements. The details of the
preparation process and a full characterization of the sample are
described elsewhere \cite{Mirko}. Here we just note that both
Auger as well as LEED measurements indicate a clean and
well-ordered surface.
For the experiments the sample was magnetized remanently along the
longitudinal axis by an external in-plane magnetic field.\\

\begin{figure}[t]
\begin{center}
\includegraphics[width=8cm,keepaspectratio=true]{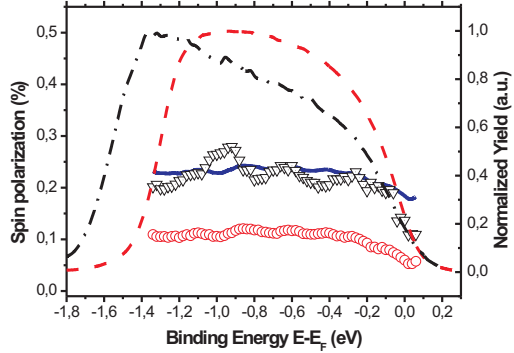}
\caption{SP-1PPE (red circles), SP-2PPE (black triangles) and
simulated SP-2PPE data (blue full line). The dashed and dotted lines
show the corresponding spin integrated 1PPE and 2PPE spectra,
respectively.} \label{SPES}
\end{center}
\end{figure}

The SP-1PPE spectra obtained in normal emission geometry show the
longitudinal in-plane SSP component and are displayed in fig.
\ref{SPES}, together with the corresponding electron yield. The
energy values in both spectra refer to the same ground state
$\ket{g}$ of the electron. Intermediate state energies can be
obtained by adding the photon energy ($3.1\,eV$) to the ground
state energy.
The  spin integrated spectra do not show any prominent features,
over an energy range down to $1.4\,eV$ below $E_F$. In contrast to
the electron yield data, the SSP data show a more structured
behavior, indicating the spin dependence of the band structure in
the magnetic compound. The spin polarization values for both methods
are almost constant for lower energies and decay both to
approximately half of their low energy value when approaching
the Fermi level. 

Comparing the SSP spectra for SP-1PPE and SP-2PPE, we observe a
general similarity. However, the spin polarization for SP-1PPE is
lowered by a factor of two compared to the SP-2PPE data. Similar
effects were obtained for the itinerant ferromagnet iron and could
be explained by a spin filter effect for hot electrons
\cite{Aeschlimann97,Aeschlimann98}.
This effect is caused by spin dependent electron relaxation
processes in the intermediate state $\ket{i}$, which can take place
at timescales comparable or even smaller than the temporal width of
the probing laser pulse. Since, in ferromagnetic materials, the
minority (spin down) electrons decay faster than the majority ones,
the resulting intermediate state polarization increases even within
a single laser pulse.

\begin{figure}[t]
\begin{center}
\includegraphics[width=7cm,keepaspectratio=true]{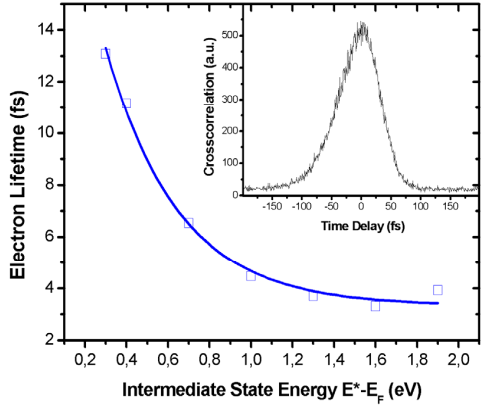}
\caption{Spin integrated electron lifetimes (blue squares) and
exponential fit (blue line) for intermediate states at $E^*$
($E_g=E^*-3.1\,eV$). Inset: example bichromatic STR-2PPE cross
correlation data used to extract $n_{\uparrow,\downarrow}.$ }
\label{tau}
\end{center}
\end{figure}

In order to quantify the impact of the spin filter effect on the
increase of polarization in the SP-2PPE data we developed a simple
numerical model containing the basic ingredients of our experiment.
First, we calculate the polarization in an intermediate state by
evaluating the time dependent numbers of up and down electrons in a
rate equation model:
\begin{equation}
\dot{n}_{\uparrow,\downarrow}=D_{\uparrow,\downarrow}p\,(t)-n_{\uparrow,\downarrow}/\tau_{\uparrow,\downarrow}
\label{eqnODE}
\end{equation}
Here $n_{\uparrow,\downarrow}$ is the occupation number for majority
and minority electrons, $\tau_{\uparrow,\downarrow}$ the spin
dependent electron lifetimes in the intermediate state, $p(t)$ a
gaussian shaped pulse with given \emph{full width at half maximum}
FWHM and $D_{\uparrow,\downarrow}$ a density of states type factor
taking care for the different spin up and spin down density of
states in the ground state. It can be
obtained 
from the measured SP-1PPE data points by setting the ratio of
$D_\uparrow$ and $D_\downarrow$ such that the polarization value
of the measurement is reproduced. The spin integrated lifetimes
$\tau_{\uparrow,\downarrow}$ were measured with time-resolved
2PPE, an experimental technique described in detail in
\cite{Aeschlimann97}. The lifetimes are shown in fig. \ref{tau}.
The ratio of $\tau_\uparrow/\tau_\downarrow$ was determined in a
bichromatic spin- and time resolved experiment (STR-2PPE), revealing
an almost energy independent ratio of $1.3$. Since in our case the
photons of a single pulse are used, the final SP-2PPE spectra can be
calculated from the solution of (\ref{eqnODE}) by simply integrating
over a second, identical pulse profile. The resulting simulated
SP-2PPE spectrum is displayed in fig. \ref{SPES} (blue full line).
One can state that the measured SP-2PPE polarization is well
reproduced, in spite of local effects caused by the particular
density of states in the intermediate state in the 2PPE process.
This is an interesting result for general interpretation of SP-1PPE
data. To draw conclusions regarding the possible use in spintronics
we want to take a closer look on the origin of the observed
electrons within the thin film.

The lifetime of an excited electron in the final state $\ket{f}$
within the crystal is very short (less than $3\,fs$), causing the
strong surface sensitivity of 1PPE data. This timescale is short
even compared to the laser FWHM. For a corresponding 2PPE process we
have to study the influence of the intermediate state at
$E^*-E_F\sim (E(\ket{f})-E_F)/2$. From standard Fermi Liquid Theory
the lifetime of this state can be estimated to be four times the one
at the final state energy. Assuming that the lifetime is
proportional to the distance the electron can travel (within the
laser width) we conclude that the mean depth of origin of electrons
emitted in a 2PPE process must be approximately four times larger
than the one for 1PPE. In other words, the possible loss in spin
polarization is not likely to occur exclusively at the very surface
but must take place deeper than roughly four times the information
depth of 1PPE, the latter corresponding to a layer of $1-2\,nm$.
Since the measured differences in SSP magnitude for the full Heusler
$Co_2FeSi$ can be explained solely by lifetime effects we conclude
that the experimentally observed loss in SSP must extend over a
range of up to $4-6\,nm$, i.e. the escape depth for 2PPE.\\


These studies were funded by the
DFG Forschergruppe FOR 559/1 "New Materials with High Spin
Polarization"


\end{document}